\documentclass[12pt]{article}
\usepackage{epsfig}

\newcommand{\be}{ \begin{equation}}
\newcommand{\ee}{ \end{equation}  }
\newcommand{\bea}{ \begin{eqnarray}}
\newcommand{\eea}{ \end{eqnarray}  }

\newcommand{\rD}{ \mbox{\rm D} }

\begin{document}

\title{Multi-boson simulation of the Schr\"odinger
functional\thanks{Talk delivered at Lattice `97, Edinburgh,
to appear in Nucl. Phys. B (Proc. Suppl.)}}

\author{Ulli Wolff\\
        Humboldt Universit\"at, 
        Institut f\"ur Physik \\ 
        Invalidenstr. 110, D-10099 Berlin, Germany}
\date{\empty}
       
\maketitle

\begin{abstract}
We discuss the choice of parameters and report some results
for unquenched simulations
of the  Schr\"odinger functional
with a non-hermitean variant of L\"uscher's multi-boson
algorithm.
\end{abstract}


Today Hybrid Monte Carlo (HMC) is the standard algorithm employed
for simulations with dynamical fermions. In spite of its general success
it seems desirable to have other methods at one's disposal.
In particular the multi-boson technique proposed by L\"uscher
\cite{MB}
seems interesting. Apart from its theoretical appeal one may
perform consistency checks and
hope for better efficiency, 
in particular with regard to slow topological
modes \cite{Topmodes}. Better numerical stability and more
flexibility in the treatment of statistical problems with
exceptional configurations \cite{JansenPHMC} may be further advantages.
Soon after L\"uscher's proposal a non-hermitean variant of
the algorithm has been advocated \cite{NHMB} and initial tests
have been performed \cite{testNHMB}, which we extend here.
Experiments with the original proposal are reported in \cite{beat}.

The contribution to the QCD Boltzmann factor from 
two flavors of dynamical quarks
is given by
\be
\exp[-S_{\rm quarks}(U)] = \det(M)^2 = |\det(M)|^2,
\ee
where $M=M(U)$ is the (sofar unimproved) Wilson Dirac operator with
the hermiticity property $M^\dag = \gamma_5 M \gamma_5$.
For the multi-boson algorithm we employ a polynomial $P(M)$
which, over the spectrum of $M$, approximates the inverse,
\be
|\det(M)|^2 = |\det P(M) |^{-2} |\det (1-R(M))|^2,
\ee
such that $R$ is a small remainder.
This enables us to represent the dominant part of $S_{\rm quarks}$
as a bosonic path integral
\bea
&\hspace{-4.0ex}&|\det P(M) |^{-2} =  \nonumber\\
&\hspace{-4.0ex}&\int \rD \phi \, \rD \phi^\dag
\exp\Biggl\{ 
-\sum_{k x} |(M-z_k)\phi_k(x)|^2 
\Biggr\},
\label{boson_action}
\eea
where $z_k, k=1,\ldots,n$, are the roots of $P$.
We now update by a sequence of
\begin{itemize}
\item 
some proposal $(U,\phi) \to (U',\phi')$ obeying
detailed balance with respect to the sum of (\ref{boson_action})
and the gluon action
\item
acceptance with probability $q(R,R',\chi)$,
\end{itemize}
where $R, R'$ are the old and new remainders and $\chi$
is a complex random field governed by
some probability distribution $\rho(\chi)$,
in our case $\rho \propto \exp(-\chi^\dag\chi)$.
This compound can be proved to be a valid algorithm if 
\be
\frac{\langle q(R,R',\chi) \rangle_\chi}{\langle q(R',R,\chi) \rangle_\chi}
= |\det(W)|^{-2}
\ee
with $W = (1-R')^{-1}(1-R)$
holds stochastically (i.e. averaged over $\chi$).
A simple (non-stochastic) solution would be
\be
q_0 = \min(1,|\det W|^{-2}).
\label{accq0}
\ee
It requires the computation
of the det of $W$.
We here use the 
``noisy algorithm'' of \cite{NHMB} corresponding to
\be
q = \min\left(1,\frac{\rho(W \chi)}{\rho(\chi)}\right).
\label{accq}
\ee
To evaluate $q$ the application of $W$ to vectors suffices,
and the required inversion of $1-R'$ with some inverter
like BiCGstab is rather uncritical.
For completeness we mention that
the variant called ``non-noisy'' in \cite{testNHMB} was found
incorrect in the implementation described there.

Following \cite{NHMB} we construct $P$ by using
Chebyshev polynomials for $R$.
On families of nested ellipses with centers at $d$,
\be
z(\theta,\varphi) = d - e \cosh(\theta + i \varphi),
\ee
they approximate the inverse
with a rate $|R| \le \exp(-(n+1)(\theta_0-\theta))$.
Here $d,e$ are fixed parameters, $\theta$ labels the 
ellipses and $\varphi\in [0,2\pi)$
traces them. The polynomial is determined (up to a factor)
by the roots $z_k=z(\theta_0,2\pi k/(n+1)), k=1,2,\ldots,n$,
lying on the ellipse
passing through the origin,
$e \cosh \theta_0 = d$. Due to even-odd symmetry, the spectrum of $M$
is symmetric under $\lambda \to 2 - \lambda$ and we hence set $d=1$.

\begin{figure}[htb]
\epsfig{file=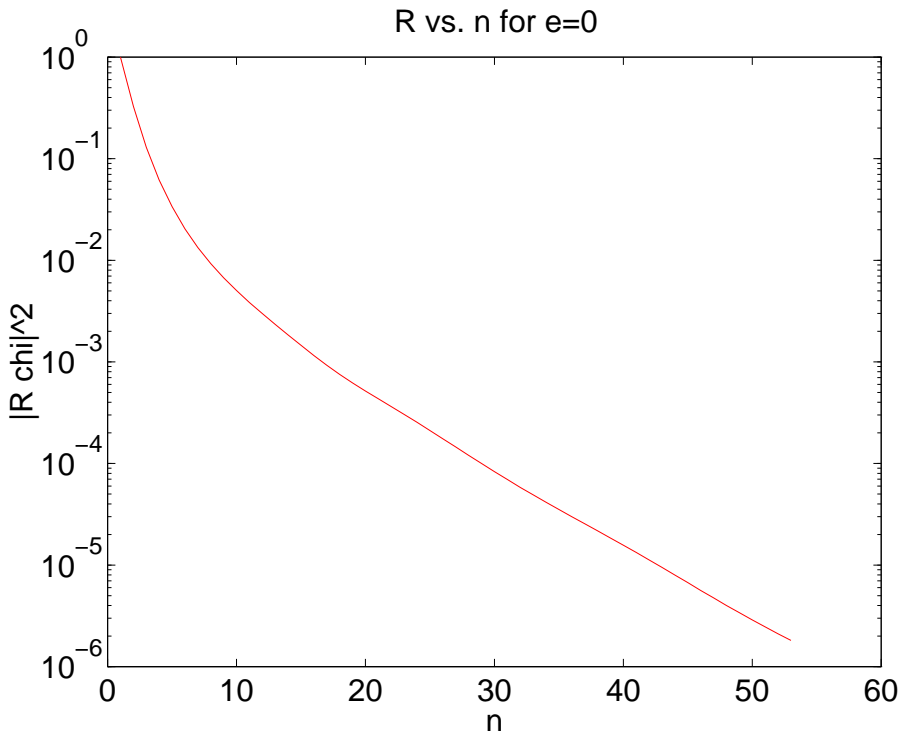,%
width=12cm}
\caption{Remainder of the inversion with $e=0$
for the Schr\"odinger functional,
 $L=T=4, \beta=6.4, K=0.15, \theta=\pi/5$, background field ``A''[9]}
\label{fig1}
\end{figure}

\begin{figure}[htb]
\epsfig{file=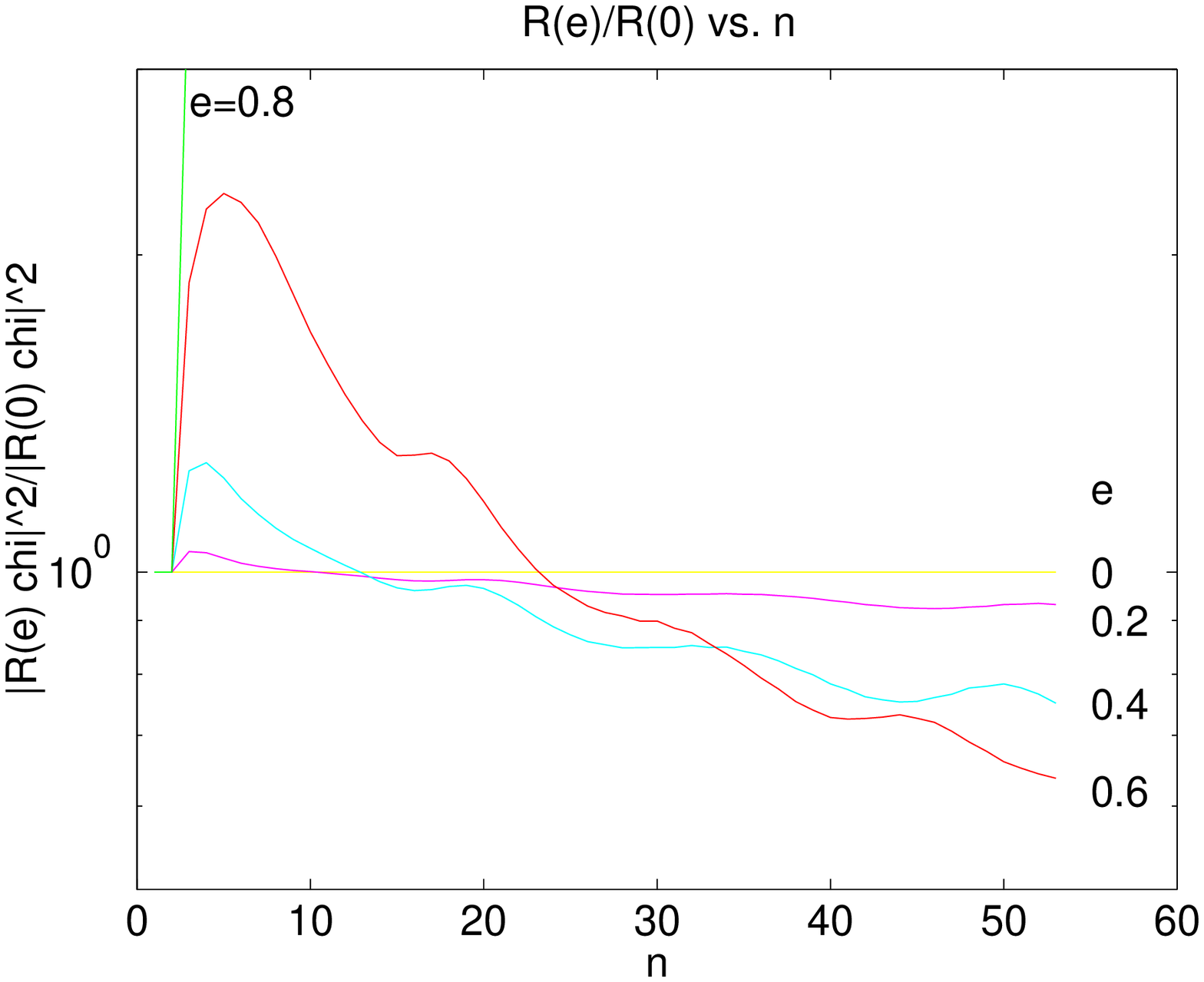,%
width=12cm}
\caption{As Fig.1, ratio of remainders for several $e$-values}
\label{fig2}
\end{figure}

To implement the correction step we have to evaluate $R\chi$.
The factorized form, $1-R =c_n M\prod (M-z_k)$,
tends to be numerically unstable \cite{JansenPHMC,elser}.
Here it can be avoided and replaced by a uniformly stable two step recursion
starting from $\chi_0=\chi$ and leading to $\chi_{n+1}=R\chi$.
It is straightforwardly
based on the standard recurrence for Chebyshev polynomials.
The intermediate $\chi_k$ are the remainders for lower degree polynomials.
We follow the recursion to investigate the choice of degree $n$ and the
focal distance $e$. With some trial parameters we produced some
equilibrated $U$-configurations, and for one of them 
Figs.\ref{fig1},\ref{fig2}
show the quality of approximation.
We see that asymptotically the best inversion is achieved for $e\sim 0.6$
which implies an oblate spectrum.
For Monte Carlo application, however, $n\sim 20$ turned out to lead to
about optimal results. In this range the value of $e$ is 
rather uncritical,
and $e=0$, where the ellipses degenerate to circles
($e\cosh\theta$ held fixed) and the polynomial to the geometric series, 
is an acceptable choice.
This is also confirmed by some direct Monte Carlo runs.
In summary, we found it practical to use inversion
as a tool to infer the spectral information necessary to
determine the parameters for simulation. The emerging 
picture was
stable for various gauge fields and random $\chi$ that were tested.

Under even-odd preconditioning one replaces $M$ by 
$\hat{M}=1-M_{\rm oe}M_{\rm eo}$ with $\det M = \det \hat{M}$, and
$M_{\rm oe}, M_{\rm eo}$ are blocks of $M$ connecting the even/odd
sublattices. An application of $\hat{M}$ has the same complexity as $M$,
but it is better conditioned. A pair of eigenvalues $\lambda, 2-\lambda$
of $M$ is mapped on one eigenvalue of $\hat{M}$ given by their product.
Under this mapping ellipses with
parameters $d=1,e$
are mapped to ellipses with $\hat{d}=1-e^2/2,\hat{e}=e^2/2$.
In this way the optimal parameters for inversion of $\hat{M}$
are given by the $e$ optimal for $M$. It again turns out, for the lattice
parameters of Figs.\ref{fig1},\ref{fig2}, that for the $n$ relevant
for efficient simulation, $\hat{e}=0$ is close to optimal.
The errors for both cases (same degree n) are connected as
$|\hat{R}| \sim |R|^2$,
which implies a much improved approximation for $\hat{M}$.
It is interesting that one can prove the relations (for even $n$)
\bea
&& \det(\hat{M}-\hat{z}_k) = \det(M-z_{k'})\\
&& \det(1-\hat{R}) = {\rm const}\times \det(1-R)
\label{acchat}
\eea
where $k'$ is some permutation of $k$.
Although $\hat{R}$ is much smaller than $R$, we get
(for every single $U$-field) the same weight from the boson fields.
As observed in \cite{testNHMB} this allows us to stick to $M$ 
for the boson fields,
which yields a much simpler structure for their 
influence on $U$-updating.
Would we use (\ref{accq0}) for the acceptance step, then
also this would be identical
for $R$ or $\hat{R}$.
In the stochastic case with (\ref{accq}), however,
preconditioning 
dramatically raises the acceptance such that $n$ may
effectively be about halved. This is due to reduced fluctuations
in $\hat{R}$ as compared to $R$.
It is trivial to derive the inequality
\be
\langle q(R,R',\chi) \rangle_{\chi} \le 
\langle  q_0(R,R')  \rangle
\ee
and its preconditioned analog. One may thus estimate the loss in acceptance
from the noisy method which turned out to be tolerable in the
preconditioned data given below (49\% down from 75\% with $q_0$
at $n=8$).
\begin{figure}
\vspace*{-2cm}
\begin{center}
\epsfig{file=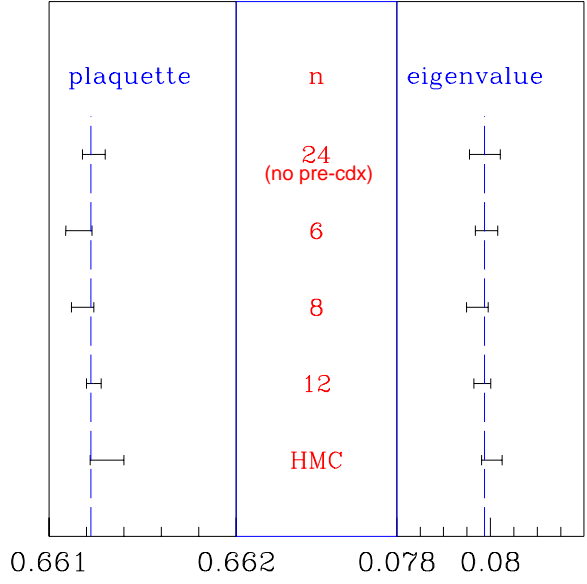,%
width=10cm}
\end{center}
\vspace*{-2cm}
\caption{Results for plaquette and lowest eigenvalue of 
$\hat{M}^\dag \hat{M}$}
\label{fig3}
\end{figure}

In Fig.\ref{fig3} results of several multi-boson simulations are shown
together with a result from preconditioned HMC for the same
parameters\cite{jansen}. 
They are obviously completely consistent for a range of acceptances
with and without preconditioning. The autocorrelation times
are given in the table.
\begin{table}
\begin{center}
\begin{tabular}{c c c c c}
n & pre & acc.(\%) & $\tau_{\rm pl}$ & $\tau_{\rm ev}$ \\[1ex]
24 & - & 88 & 4.0 & 21 \\[.5ex]
6 & x &  27 & 1.9 & 3.7 \\
8 & x &  49 & 1.3 & 3.1 \\
12 & x & 77 & 1.3 & 3.9 \\
HMC& x &    & 1.5 & 1.5  
\end{tabular}
\end{center}
\end{table}
All $\tau$ refer to units of 1000 $M\phi$ applications.
The proposals are generated with a certain combination
of microcanonical and heatbath sweeps.
While the multi-boson algorithm seems advantageous for the
plaquette, there is an advantage 
to HMC
for the eigenvalue. In actual CPU time on the Quadrics Q1
the multi-boson algorithm is faster for both quantities
for our particular implementations.
We plan to clarify this issue further
by a simulation on an $8^4$ lattice, but it seems likely that without further
new ideas there are no large factors in efficiency attainable
between the two rather different methods.

I would like to thank Burkhard Bunk for discussions.


\end{document}